\documentclass[conference]{IEEEtran}
\IEEEoverridecommandlockouts
\usepackage{cite}
\usepackage{amsmath,amssymb,amsfonts}
\usepackage{algorithmic}
\usepackage{graphicx}
\usepackage{textcomp}
\usepackage{xcolor}
\usepackage{amssymb}
\usepackage{booktabs} 
\usepackage[a4paper, total={184mm,239mm}]{geometry}
\usepackage{algorithm}
\usepackage{algorithmic}
\usepackage{array}
\usepackage{booktabs}
\usepackage[colorlinks=true,linkcolor=black,urlcolor=magenta,citecolor=black]{hyperref}

\def\BibTeX{{\rm B\kern-.05em{\sc i\kern-.025em b}\kern-.08em
    T\kern-.1667em\lower.7ex\hbox{E}\kern-.125emX}}
\begin{document}

\title{NDFT: Accelerating Density Functional Theory Calculations via Hardware/Software Co-Design on Near-Data Computing System} 

\newcommand{\summary}[1]{\textcolor{blue}{#1}}
\newcommand{\confused}[1]{\textcolor{red}{#1}}
\newcommand{\todo}[1]{\textcolor{magenta}{#1}}
\newcommand{\jiang}[1]{\textcolor{purple}{#1}}



\author{\IEEEauthorblockN{Qingcai Jiang\textsuperscript{1}$^*$, Buxin Tu\textsuperscript{1}$^*$, Xiaoyu Hao\textsuperscript{1}, Junshi Chen\textsuperscript{1},\textsuperscript{2} and Hong An\textsuperscript{1},\textsuperscript{2},\textsuperscript{\dag}}
\IEEEauthorblockA{\textsuperscript{1}School of Computer Science and Technology, University of Science and Technology of China, Hefei, China \\
}
\IEEEauthorblockA{\textsuperscript{2}Laoshan Laboratory, Qingdao, China \\
}

}

\maketitle

\def\thefootnote{*}\footnotetext{Equal contribution, listed alphabetically}
\def\thefootnote{\textsuperscript{\dag}}\footnotetext{Corresponding Author}

\begin{abstract}
Linear-response time-dependent Density Functional Theory (LR-TDDFT) is a widely used method for accurately predicting the excited-state properties of physical systems.
Previous works have attempted to accelerate LR-TDDFT using heterogeneous systems such as GPUs, FPGAs, and the Sunway architecture.
However, a major drawback of these approaches is the constant data movement between host memory and the memory of the heterogeneous systems, which results in substantial \textit{data movement overhead}.
Moreover, these works focus primarily on optimizing the compute-intensive portions of LR-TDDFT, despite the fact that the calculation steps are fundamentally \textit{memory-bound}.

To address these challenges, we propose NDFT, a \underline{N}ear-\underline{D}ata Density \underline{F}unctional \underline{T}heory framework.
Specifically, we design a novel task partitioning and scheduling mechanism to offload each part of LR-TDDFT to the most suitable computing units within a CPU-NDP system.
Additionally, we implement a hardware/software co-optimization of a critical kernel in LR-TDDFT to further enhance performance on the CPU-NDP system.
Our results show that NDFT achieves performance improvements of 5.2x and 2.5x over CPU and GPU baselines, respectively, on a large physical system. 

\end{abstract}

\begin{IEEEkeywords}
Near-Data Processing,
Density Functional Theory,
Workload Scheduling,
Hardware/Software Co-Design
\end{IEEEkeywords}

\section{Introduction}
First-principle calculations, also known as \textit{ab initio} methods, predict the properties of materials by solving the electronic motion equations derived from the fundamental principles of quantum mechanics.
Among these methods, density functional theory (DFT) is widely used to determine the ground-state properties of systems by focusing on electron density rather than wavefunctions.
Building on DFT, \textbf{linear response time-dependent density functional theory (LR-TDDFT)} is particularly notable for its ability to accurately describe the excited states of complex systems.
LR-TDDFT is extensively applied in material analysis, condensed matter physics, and quantum chemistry.
However, it often encounters significant computational challenges, including high execution time, memory complexity, and frequent data accesses, which can easily lead to memory bottlenecks.
Given the crucial role of LR-TDDFT in scientific research and the computational challenges it presents, exploring optimization techniques for this method has become a prominent topic in the field of high-performance computing.

Recent studies have tried to accelerate LR-TDDFT calculations using heterogeneous systems such as graphics processing units (GPUs)~\cite{jia2019parallel} and the Sunway architecture~\cite{jiang2024extending}. 
These architectures leverage parallel processing capabilities to execute a large number of computations simultaneously, significantly speeding up the \textit{compute-intensive} aspects of the software.
However, executing LR-TDDFT on these systems requires frequent data transfers between main memory and local memory.
This overhead from this data movement can negate the performance gains achieved by heterogeneous architectures, leading to a phenomenon known as the data movement bottleneck.

\textbf{Near-data processing (NDP)} is a promising solution to this bottleneck in modern computing systems.
By positioning computing units closer to the main memory, the NDP mechanism reduces the overhead associated with data access~\cite{mutlu2022modern}.
With recent advancements in memory technology~\cite{aga2019co,gu2020ipim,ahn2015scalable,boroumand2018google,farmahini2015nda,asgari2021fafnir,asghari2016chameleon,dai2022dimmining,feng2022menda} and the growing demand from data-intensive applications~\cite{chen2014data,boroumand2018google}, the near-memory computing paradigm has become increasingly feasible~\cite{hu2018stacking,jeddeloh2012hybrid}. 


However, effectively harnessing the potential of NDP systems to accelerate LR-TDDFT calculations presents several key challenges.
First, NDP systems typically operate alongside CPU systems, forming a CPU-NDP system.
For LR-TDDFT calculations, which involve both compute-intensive and memory-intensive kernels, such systems require a precise scheduling and partitioning scheme to ensure efficient execution.
Second, in the pseudopotential computation of LR-TDDFT (see Section~\ref{sec:analysis_pseudopotential}), each process stores its copy of the data in memory, leading to significant redundancy given the large number of NDP cores and corresponding processes. 
Therefore, we need to rethink the data organization of pseudopotential in the NDP scenario to mitigate this issue. 



To address these challenges, \textbf{our goal} in this work is to design a partitioning and scheduling mechanism specifically tailored for DFT calculations to achieve better performance, along with a hardware/software co-design solution that optimizes the data organization and memory access in the pseudopotential calculations. 

This paper makes the following key contributions:
\begin{itemize}
    \item We conduct a detailed study of the computation and memory characteristics in the LR-TDDFT application, identifying the key performance bottlenecks in its execution and determining LR-TDDFT as a promising application for NDP systems. 
    \item We design NDFT, a near-data density functional theory calculation framework, featuring a novel workload partitioning and scheduling mechanism, along with a hardware/software co-design based on a CPU-NDP heterogeneous architecture. 
    \item We evaluate NDFT across physical systems of varying sizes and compare it with CPU and GPU baselines. The results show that NDFT achieves a performance speedup of 5.2x and 2.5x over CPU and GPU baselines on a large physical system. 
\end{itemize}

\section{Background}
\subsection{LR-TDDFT}
\label{sec:lrtddft-background}
\begin{figure}[htbp]
    \centering
    \vspace{-4mm}
    \includegraphics[width=0.8\linewidth]{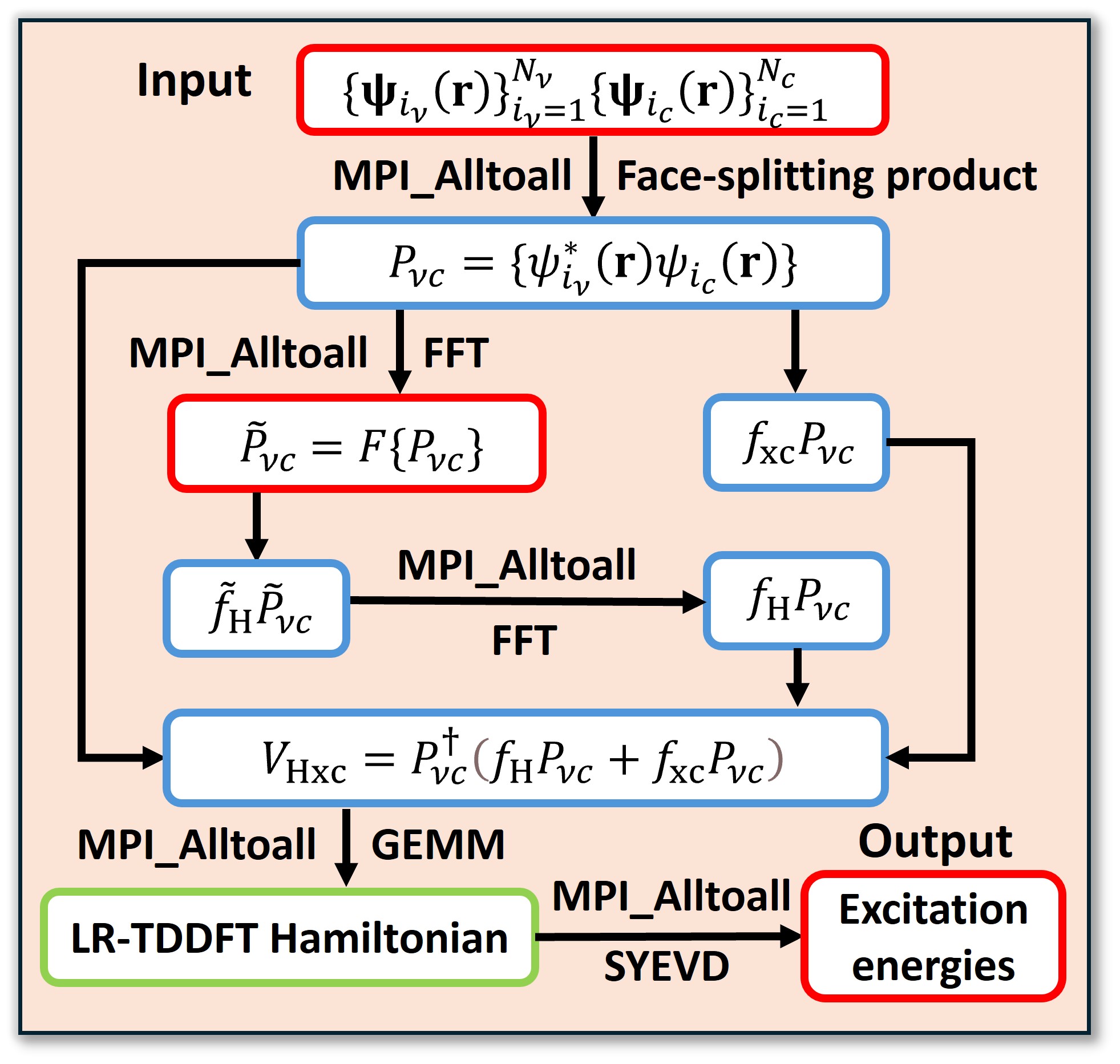}\\
    \caption{Computation flowchart of LR-TDDFT. $\psi_{i_{v}}(\mathbf{r}) \text { and } \psi_{i_{c}}(\mathbf{r})$ stand for the valence and conduction orbitals in real space ($\left\{\mathbf{r}_{\mathbf{i}}\right\}_{i=1}^{N_{r}}$.}
    \label{fig:lrtddft-flow}
    \vspace{-4mm}
\end{figure}
LR-TDDFT~\cite{runge1984density} is a widely used computational method for studying the excited-state properties of molecules and solids in condensed matter physics, computational chemistry, and materials science.
In recent years, advancements in both computational hardware and numerical algorithms have significantly improved the scalability and performance of LR-TDDFT, enabling its application to large and complex physical systems~\cite{wan2021hybrid,liu2011analytical}.
As shown in Figure~\ref{fig:lrtddft-flow}, the computation of LR-TDDFT involves several major operations: (1) fast Fourier transform (\textbf{FFT}), (2) face-splitting product, i.e., \textbf{point-to-point multiplication} between two vectors, (3) general matrix multiplications (\textbf{GEMM}), (4) \textbf{MPI\_Alltoall} between all processes to transpose data, and (5) diagonalization (\textbf{SYEVD}).

Due to its scientific significance, numerous prior works have focused on accelerating the \textit{computation-intensive} aspects of LR-TDDFT using heterogeneous architectures, such as GPUs~\cite{jia2019parallel, wang2011large}, the Sunway supercomputer~\cite{jiang2024extending, jiang2024pwdft}, and field-programmable gate arrays (FPGAs)~\cite{ramaswami2020efficient}.
However, accelerating LR-TDDFT on these heterogeneous architectures presents two major problems: (1) These approaches primarily target the \textit{compute-intensive} components, neglecting the \textit{memory-intensive} parts, which constitute a significant portion of LR-TDDFT's overall execution time; (2) Executing kernels on heterogeneous architectures requires frequent data transfers between the host device's memory and the memory of the heterogeneous device, resulting in substantial bottlenecks.


\subsection{NDP Architecture}
The fundamental idea of NDP is to place computing logic close to where data resides, offering significant potential for improving the performance of \textit{memory-intensive applications}.
NDP can be implemented by integrating processing logic within DRAM banks~\cite{aga2019co,gu2020ipim}, in the logic layer of 3D-stacked memory~\cite{ahn2015scalable,boroumand2018google,farmahini2015nda}, or within the buffer chip of DDR-DIMMs~\cite{asgari2021fafnir,asghari2016chameleon,dai2022dimmining,feng2022menda}. 
\begin{figure}[htbp]
    \centering
    \vspace{-4mm}
    \includegraphics[width=0.8\linewidth]{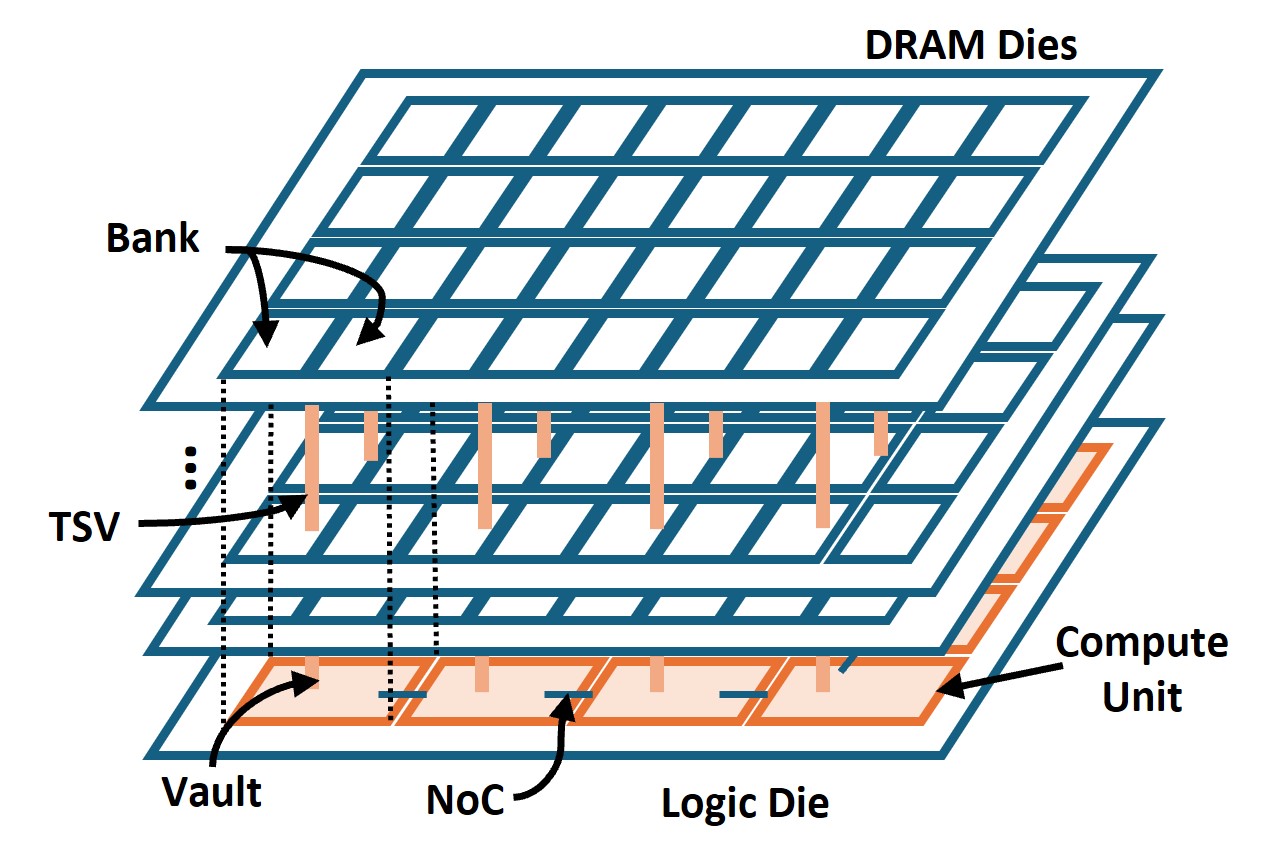}\\
    \vspace{-4mm}
    \caption{An example of 3D-stacked memory.}
    \label{fig:3d-stacked}
\end{figure}
As shown in Figure~\ref{fig:3d-stacked}, in such architectures, the bottom-most layer, known as the logic layer, can be equipped with \textbf{wimpy cores} that perform simple, parallelizable tasks close to the data. 
These cores leverage the high internal bandwidth and low data movement latency of the memory stacks to improve efficiency. 
Typical 3D-stacked NDP systems, such as Hybrid Memory Cube (HMC)~\cite{hmc} and High Bandwidth Memory (HBM)~\cite{hbm,hbm2}, consist of multiple memory stacks interconnected in a memory network.
By bringing computation closer to the data, NDP systems take advantage of the high bandwidth within the memory structure and reduce data access latency. 

\subsection{CPU-NDP Hybrid Execution}
\begin{figure}[htbp]
    \centering
    \vspace{-4mm}
    \includegraphics[width=\linewidth]{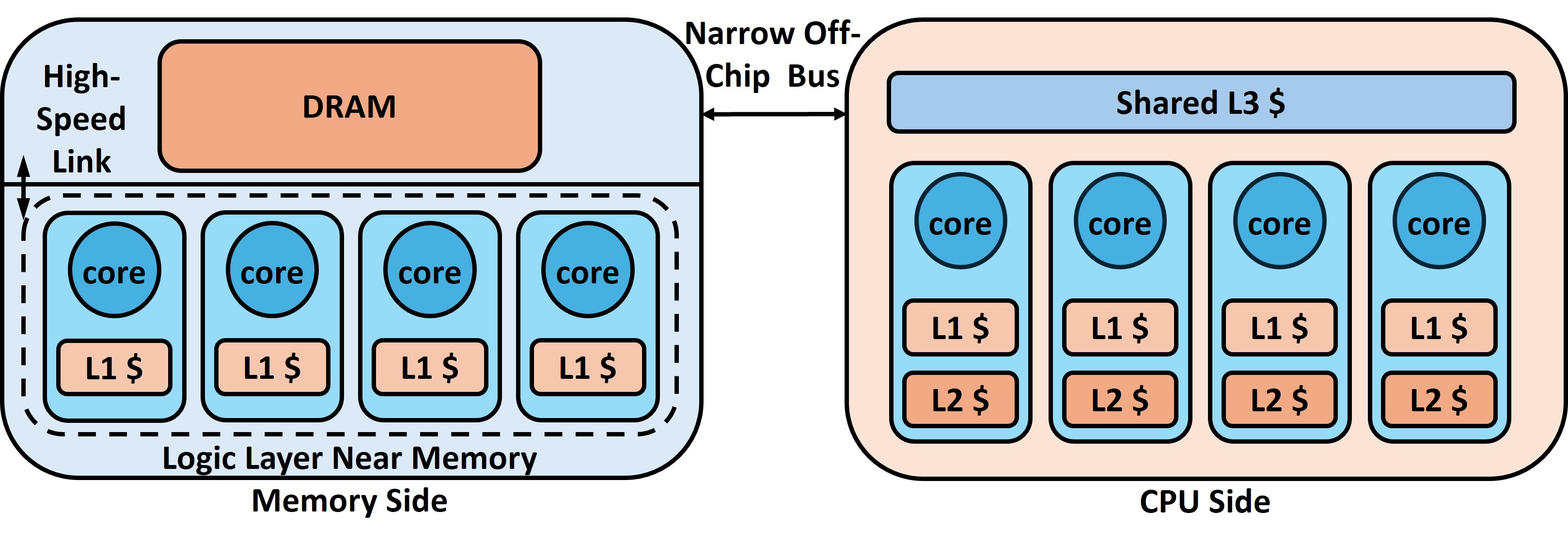}\\
    \vspace{-4mm}
    \caption{A high-level CPU-NDP architecture.}
    \label{fig:cpu-ndp}
\end{figure}

In a typical configuration, NDP cores operate alongside CPU cores within a computing system, forming what we refer to as the CPU-NDP architecture in this paper, as illustrated in Figure~\ref{fig:cpu-ndp}. 
This architecture includes two types of computing units: the CPU, which excels at complex, compute-intensive tasks but suffers from high-latency and low-bandwidth memory access, and the NDP cores, which are more numerous and benefit from lower-latency, higher-bandwidth memory access, but feature simpler logic. 
Thus, CPU cores are suitable for \textit{compute-intensive} workloads, while NDP cores handle \textit{memory-intensive} tasks more efficiently.
In such systems, a key challenge is determining which code regions of a program should be offloaded to NDP for optimal performance. 
Several prior works have attempted to solve this problem~\cite{wei2022pimprof,ghiasi2022alp,jiang20243}. 
However, none of these approaches have been applied to exploit the performance potential of DFT applications in CPU-NDP systems. 

\section{Motivational Study}

In this section, we conduct a detailed study on several key characteristics of LR-TDDFT, aiming to identify the key bottlenecks in CPU-NDP hybrid execution and demonstrate its potential for performance improvement on CPU-NDP heterogeneous architecture. 
\subsection{LR-TDDFT kernel characteristics}
\begin{figure}[htbp]
    \centering
    \vspace{-3mm}
    \includegraphics[width=\linewidth]{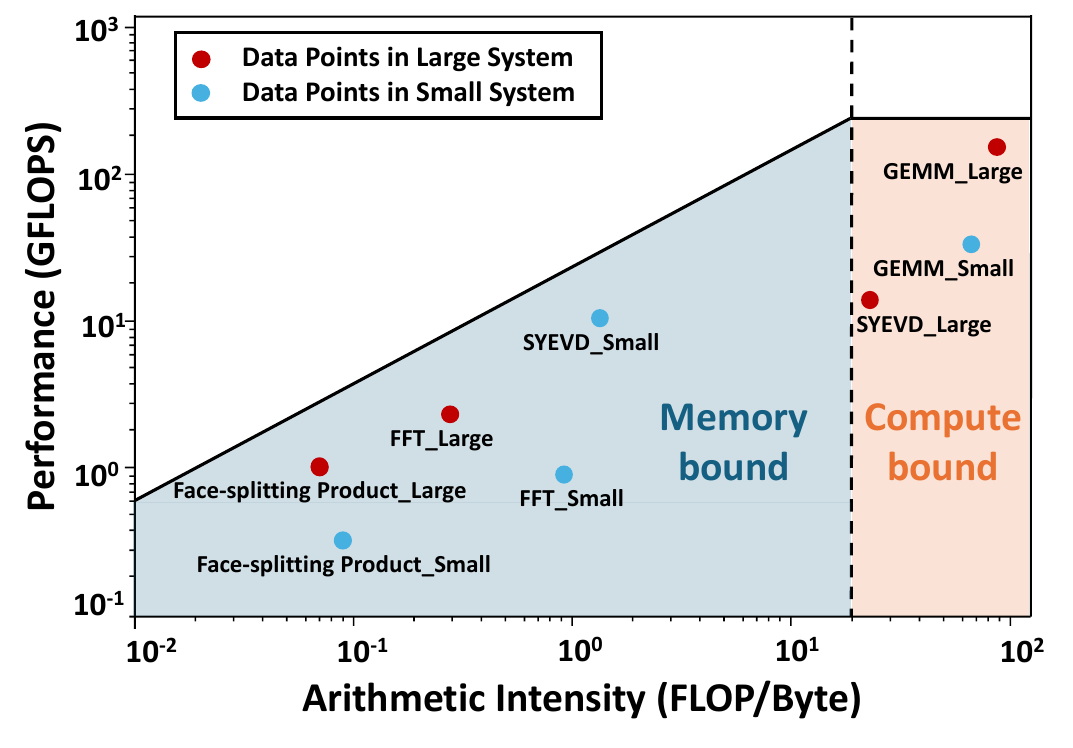}
    \vspace{-3mm}
    \caption{Roofline model analysis of LR-TDDFT kernels across two different system sizes.}
    \label{fig:roofline}
    \vspace{-3mm}
\end{figure}
We analyze the performance characteristics of representative kernels in LR-TDDFT calculations on the Roofline model, as introduced in Section~\ref{sec:lrtddft-background}, using two different physical system sizes: Si\_64 (small system) and Si\_1024 (large system). 
The Roofline model~\cite{williams2009roofline} integrates floating-point operations (FLOPs) and arithmetic intensity (AI) to assess the computational and memory characteristics of a given kernel.
We use Intel VTune Profiler~\cite{intel_vtune} for the analysis, with system configurations described in Section~\ref{sec:methodology}.
From the results shown in Figure~\ref{fig:roofline}, we make three key observations: 
(1) LR-TDDFT calculations are fundamentally \textbf{memory-bound}: most computing kernels in LR-TDDFT fall within the memory-bound region, and the MPI\_Alltoall operations used in LR-TDDFT introduce a large amount of irregular memory accesses~\cite{jiang2024extending}.
(2) Different kernels in LR-TDDFT calculations exhibit distinct compute-bound or memory-bound characteristics: FFT kernels are memory-bound, while GEMM kernels are compute-bound across different physical system sizes.
(3) The compute- or memory-bound nature of LR-TDDFT kernels depends on the system size: for large physical systems, the SYEVD operation is compute-bound, whereas, for small systems, SYEVD is memory-bound; GEMM becomes more compute-bound as the system size increases. 

Based on these observations, we conclude that LR-TDDFT is well-suited for CPU-NDP acceleration due to its diverse kernel characteristics.
To fully leverage the strengths of both CPU and NDP cores, the system requires an intelligent scheduling mechanism that can dynamically allocate tasks to the appropriate computing units based on their characteristics.
For example, memory-bound kernels such as FFT could benefit from execution on NDP units optimized for memory access, while compute-bound kernels like GEMM and SYEVD (in large systems) would perform better on CPU cores, which excel at handling complex computations.
However, designing a precise task scheduling framework for LR-TDDFT in a CPU-NDP system is not trivial, as it faces two main challenges:
(1) It must account not only for the suitability between kernels and computing units but also for the \textit{data movement overhead} between the two types of computing units. 
The data required by the kernels must also be scheduled accordingly to the appropriate execution units.
(2) The selection of offloading granularity, ranging from fine-grained instructions to basic blocks or entire kernels, requires careful consideration.


\subsection{Analysis of pseudopotential calculations}
\label{sec:analysis_pseudopotential}

Pseudopotentials are commonly used in DFT calculations to simplify the treatment of inner electrons, as the outer valence electrons primarily determine an atom's physical and chemical properties.
In LR-TDDFT, pseudopotentials simulate the influence of the atomic nucleus and inner electrons.
During this procedure, each wavefunction is updated based on the pseudopotential information associated with each atom.
In a multi-process execution, each process must maintain a complete copy of the pseudopotential information in its memory space.
However, in systems with numerous cores, such as NDP architectures~\cite{tian2023abndp,ghiasi2022alp}, these redundant data copies can lead to significant memory inefficiency.

\begin{table}[htbp]
\caption{Memory Footprint of Pseudopotentials in CPU and NDP Systems}
\centering
\vspace{-2mm}
\begin{tabular}{c|c|c}
\toprule
\textbf{Memory Footprint} & \textbf{size (GB)} & \textbf{Percentage (\%)} \\
\midrule
NDP in Small system & 4.43 & 6.92 \\
\midrule
CPU in Small system & 1.84 & 2.88 \\
\midrule
NDP in Large system & 35.3 & 55.15 \\
\midrule
CPU in Large system & 13.8 & 21.56 \\
\bottomrule
\end{tabular}
\vspace{-3mm}
\label{tab:memory-usage}
\end{table}
We profile the memory footprint during LR-TDDFT calculations on typical NDP and CPU systems using two physical systems of different scales, where the NDP system and the CPU system are isolated. 
The configuration is detailed in Section~\ref{sec:methodology}. 
As shown in Table~\ref{tab:memory-usage}, our observations are as follows: (1) In the small physical system, the memory footprint of pseudopotentials in NDP execution occupies 6.92\% of the system memory, which is 140.2\% higher than that in CPU execution; (2) In the large physical system, the memory footprint reaches 55.15\%, representing an increase of 155.7\% compared to CPU execution. 
We conclude that the large number of NDP cores leads to an increase in parallel processes, resulting in pseudopotentials consuming a significant amount of memory space.
Moreover, in more complex physical systems, the traditional per-process pseudopotential approach causes an out-of-memory (OOM) problem. 
Therefore, optimizing the data structure of pseudopotential information in memory is critical to reducing the substantial memory footprint in the CPU-NDP architecture.


\section{NDFT Design}
In this section, we elaborate on the design of NDFT, aiming to maximize the performance potential of the CPU-NDP system for DFT calculations. 
This includes a novel workload partitioning and scheduling mechanism, which is introduced in Section~\ref{sec:worklod_scheduling}, as well as a hardware/software co-design strategy to better orchestrate an important computational scheme in LR-TDDFT calculations, which is covered in Section~\ref{sec:opt_pseudo_alg} and~\ref{sec:shared-memory}, respectively.

\subsection{Workload Partition and Scheduling}
\label{sec:worklod_scheduling}
\subsubsection{Offload Granularity}
To provide a better offloading granularity for LR-TDDFT, it is crucial to understand the costs associated with scheduling contiguous code segments to different computing units.
We identify two primary sources of overhead:
(1) The \textit{data transfer overhead}, which arises from maintaining the consistency of data across two computing units. This overhead is proportional to the amount of data that needs to be transferred.
(2) The \textit{context switch overhead}, which involves synchronizing the context (e.g., register values) between threads on different units, is a constant time cost. 

In this work, we choose \textit{function-level} offloading granularity based on two key observations:
(1) Dividing the program into too many small segments introduces significant offloading overhead.
(2) Most functions in LR-TDDFT exhibit consistent compute/memory characteristics throughout their execution. For example, GEMM and FFT maintain the same computing and memory access patterns across the entire function.

\subsubsection{Cost-Aware Offloading mechanism}
To fully leverage the performance benefits of a heterogeneous CPU-NDP system, we design a kernel offloading mechanism in NDFT based on a static code analyzer (SCA)~\cite{iaca2017, lattner2004llvm}.
The SCA analyzes the intrinsic properties of a given code segment, such as estimated execution time, memory access patterns, and instruction dependencies.
These insights help inform decisions on whether each code segment is compute- or memory-bound, and provide an estimation of the data transfer overhead when scheduling different code segments across CPU and NDP units.
In this work, we leverage the SCA to profile the compute and memory intensity of each function and use this information, along with a cost model that accounts for the overhead of scheduling between CPU and NDP cores, to formulate a cost-aware offloading mechanism.

\vspace{-3mm}
\begin{equation}
\label{eq:scheduling_overhaed}
\text{Scheduling Overhead} = \sum_{i \in \text{NDP}} \sum_{j \in \text{CPU}}\left(DT(i, j) + CXT\right)
\end{equation}
\vspace{-3mm}

By evaluating the suitability of each function for execution on either CPU or NDP cores, the SCA allows us to determine whether it is more efficient to offload the function to the NDP or retain it on the CPU.
The overall scheduling strategy then incorporates the scheduling overhead, as shown in  (\ref{eq:scheduling_overhaed}), which is defined as the sum of data transfer (DT) overhead and context switch (CXT) overhead between the CPU and NDP cores.
This approach enables a cost-aware offloading mechanism that balances the performance gains from offloading with the associated overheads, optimizing the performance of the heterogeneous system.

\subsection{Optimization of Pseudopotential Algorithm}
\label{sec:opt_pseudo_alg}
 
\begin{figure}[htbp]
    \centering
    \vspace{-3mm}
    \includegraphics[width=\linewidth]{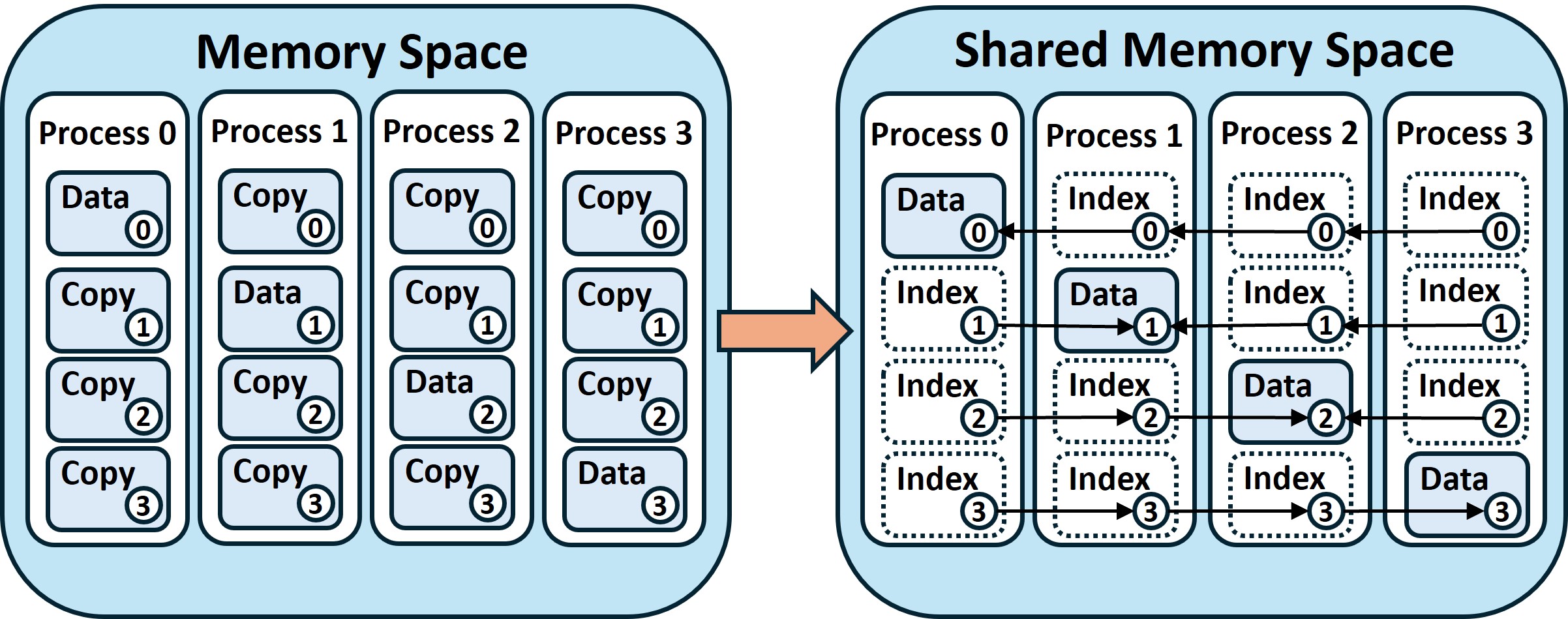}\\
    \caption{The data structure optimization to eliminate the data redundancy of pseudopotential.}
    \label{fig:data-sharing}
    \vspace{-3mm}
\end{figure}

To reduce the memory footprint in LR-TDDFT and enhance the scalability for large physical systems, we propose an optimization based on an improved pseudopotential data structure.
As shown in Figure~\ref{fig:data-sharing}, each process stores \textbf{only a portion of the pseudopotentials}, along with indices pointing to the pseudopotentials in other processes, accessing the shared data via these indices.
Compared to the original approach, our optimization enables seamless data sharing between processes, significantly reducing the memory footprint and making LR-TDDFT more suitable for our CPU-NDP heterogeneous design.

\begin{algorithm}[htbp]
 \renewcommand{\algorithmicrequire}{\textbf{Input:}}
 \renewcommand{\algorithmicensure}{\textbf{Output:}}
\caption{Pseudopotential Algorithm in LR-TDDFT, with the gray background highlighting our optimizations.}
\label{alg:psedupotential}
 \begin{algorithmic}[1]
     \REQUIRE Pseudopotential
     \ENSURE Wavefunction updated using pseudopotential
    \FOR{each local atom}
        \STATE calculate pseudopotential
    \ENDFOR
    \FOR{each global atom}
        \IF{this process holds the information}
            \FOR {each atom's pseudopotential}
                \STATE \colorbox{gray!20}{calculate the data size}
                \STATE \colorbox{gray!20}{allocate a continuous space in shared memory}
                \STATE \colorbox{gray!20}{write local pseudopotential information as}
                \colorbox{gray!20}{a block into shared memory}
            \ENDFOR
        \ELSE
            \FOR{each shared memory block}
                \STATE \colorbox{gray!20}{obtain the address of the shared block}
            \ENDFOR
        \ENDIF
    \ENDFOR
    \FOR{each local wavefunction}
        \FOR{each pseudopotential of every atom}
            \STATE \colorbox{gray!20} {access pseudopotential via shared block address}
            \STATE apply pseudopotential to the wavefunction
        \ENDFOR
    \ENDFOR
 \end{algorithmic}  
 \end{algorithm}
 
In our optimization, we reduce the memory overhead of each process by addressing the complexity of each atom’s pseudopotential data, which includes arrays of integers and double-precision floating-point matrices, through the design of a \textbf{shared block} to efficiently manage and share this information.
This data structure facilitates unified management of pseudopotential data and simplifies data sharing.
As shown in Algorithm~\ref{alg:psedupotential}, the highlighted gray section represents the main optimization of the algorithm:
(1) In lines 7-9, the algorithm reorganizes the pseudopotential information of each atom into a shared block and writes it into shared memory.
(2) In line 13, the algorithm retrieves the addresses of pseudopotential data in shared memory based on the distribution of the shared block.
(3) In line 19, the algorithm accesses the data from shared memory for wavefunction updating.

Although this approach significantly reduces memory usage, it introduces the need for inter-process data communication, leading to some performance overhead.
We design a novel shared memory hardware structure to mitigate this overhead, detailed in Section~\ref{sec:shared-memory}.

\subsection{Shared Memory Design for Pseudopotential}
\label{sec:shared-memory}

As we improve the data structure of the pseudopotential in each process, data communication is needed to synchronize the pseudopotential.
To mitigate this overhead, we design shared memory based on scratchpad memory (SPM) for processes to share the pseudopotential information within the same stack, and a hierarchical communication scheme to manage the communication between the shared memory of different stacks. 

\begin{figure}[htbp]
    \vspace{-3mm}
    \centering
    \includegraphics[width=\linewidth]{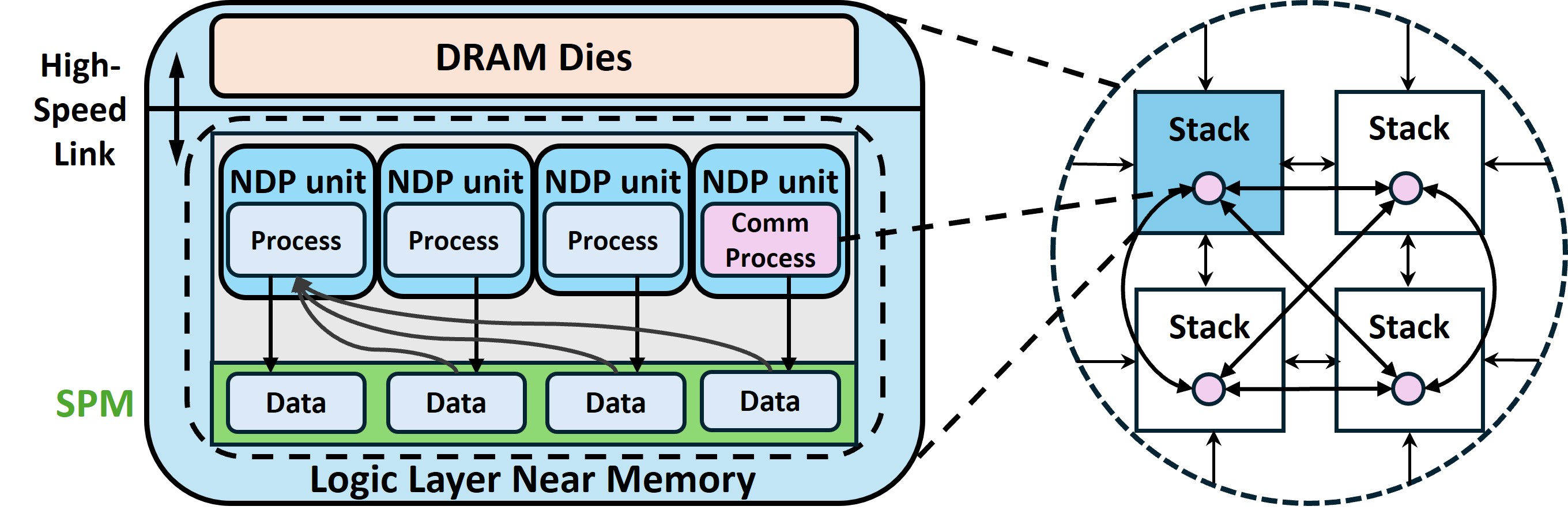}\\
    \caption{Shared Memory Hardware Design and Hierarchical Communication Scheme. Green color highlights the shared memory architecture.}
    \label{fig:shared-memory}
    \vspace{-3mm}
\end{figure}

\textbf{SPM-based Shared Memory.} The left part of Figure~\ref{fig:shared-memory} shows our shared memory hardware design.
We design an SPM in the logic layer of each memory stack to enable the shared memory functionality.
The processes within the same stack can access local pseudopotentials and also access the pseudopotentials of other processes in SPM-based shared memory. 
The high read/write speed and low access cost of SPM~\cite{banakar2002scratchpad} allow these processes to efficiently access pseudopotentials, avoiding the expensive memory access in DRAM. 

In addition, we develop a set of communication APIs to support shared memory functionality. 
As shown in Table~\ref{tab:comm-primitives}, these APIs implement multiple synchronization semantics, providing the flexibility and comprehensiveness needed for essential communication primitives.
For example, \texttt{NDFT\_Alloc\_Shared} enables NDP units to allocate memory for pseudopotential data in \texttt{sharedBL}. 
The function's parameters include the pseudopotential data (\texttt{pseu\_info}) and the NDP unit ID (\texttt{stID* id}), and it returns a \texttt{sharedBL}. 
The definition of \texttt{sharedBL} corresponds to the shared block introduced in Section~\ref{sec:opt_pseudo_alg}. 
We also design primitives such as \texttt{NDFT\_Read} and \texttt{NDFT\_Write} to facilitate reading and writing pseudopotentials within the shared block. 
By managing pseudopotentials with SPM-based shared memory, our design reduces communication overhead between NDP units within the same stack. 
To further enhance scalability, we implement a hierarchical communication scheme.

\begin{table}[htbp]
\vspace{-3mm}
    \centering
    \caption{Programming Interfaces of NDFT}
    \begin{tabular}{ll}
    \toprule
    \textbf{NDFT's Message Programming Interface(i.e., API)}& \begin{tabular}[c]{@{}c@{}}\end{tabular}
         \\ 
    \midrule
    sharedBL NDFT\_Alloc\_Shared (Var* pseu\_info, stID* id);\\
    void NDFT\_Read (sharedBL* data, void* address, int length);\\
    void NDFT\_Write (sharedBL* data, void* address, int length);\\
    void NDFT\_Read\_Remote (sharedBL* data, void* address, int length,\\ stID* source\_id, stID* dest\_id);\\
    void NDFT\_Write\_Remote (sharedBL* data, void* address, int length,\\ stID* source\_id, stID* dest\_id);\\
    void NDFT\_Broadcast (sharedBL* data);\\
    ......\\
    \bottomrule
    \end{tabular}
    \vspace{-3mm}
    \label{tab:comm-primitives}
\end{table}

\textbf{Hierarchical Communication Scheme.}
For pseudopotential data communication between different stacks, we designate one NDP unit per stack as a communication arbiter, which runs a \textit{comm process} (indicated in pink in Figure~\ref{fig:shared-memory}) to manage inter-stack communication.
When a process requests data from another stack via \texttt{NDFT\_Read\_Remote}, it submits the request to the \textit{comm process} within its own stack.
The \texttt{dest\_id} parameter specifies the stack ID from which the data is needed.
After exchanging data with the communication process of the destination stack, the local \textit{communication process} writes the data into shared memory and returns the index to the requesting process.
This hierarchical design acts as a filter, maximizing intra-stack communication and only transmitting essential data across stacks, thereby reducing overall communication overhead.


\section{Evaluation Methodology}
\label{sec:methodology}
\textbf{System model.} 
We implement NDFT in zsim~\cite{sanchez2013zsim}, a fast and accurate Pin-based simulator, and use Ramulator~\cite{kim2015ramulator} as the memory simulator. 
Table~\ref{tab:cpu-ndp-config} summarizes the NDFT's CPU-NDP system configurations.

\begin{table}[htbp]
    \centering
    \caption{CPU-NDP System Configuration}
    \begin{tabular}{cc}
    \toprule
    \textbf{CPU}    & \begin{tabular}[c]{@{}c@{}}8 General purpose core;\\         3 GHz, 4-way superscalar;\\         32 KB L1I/D, 256 KB L2, 2 MB L3\end{tabular}               \\ 
    \midrule
    \textbf{NDP}    & \begin{tabular}[c]{@{}c@{}}8 NDP units per stack, 2GHz, in order;\\ 64 GB in total, 512 MB per unit;\\ 2 GHz, 2 cores per NDP unit, 32KB L1I/D\\ Shared Memory: 16 KB per core, 256 KB per stack \end{tabular} \\
    \midrule
    \textbf{Memory} & \begin{tabular}[c]{@{}c@{}}HBM2, 4 × 4 stacks in mesh, 8 channels per stack;\\ 128-bit bus width, 1000Mhz, 64 GB capacity\end{tabular}                      \\ 
    \bottomrule
    \end{tabular}
    \label{tab:cpu-ndp-config}
    \vspace{-3mm}
\end{table}

\textbf{Baseline Architectures.} 
We compare NDFT with two baselines:
(1) CPU: 2 CPUs of Intel Xeon E5-2695@2.40 GHz, 12 cores per socket, with 64 GB DDR4 memory;
(2) GPU: 2 GPUs of NVIDIA V100 in DGX-1 server.

\textbf{Physical Systems.}
We use various crystal silicon systems with 16, 32, 64 (small system), 128, 256, 1024 (large system), and 2048 silicon atoms, denoted as Si\_16, Si\_32, Si\_64, Si\_128, Si\_256, Si\_1024, and Si\_2048, respectively.

\section{Evaluation Result}

\subsection{Performance Analysis on Execution Time}

Figure~\ref{fig:execution-time} shows the performance on CPU, GPU, and NDFT, along with the time distribution across different kernels in 2 different sizes of physical systems. 
The execution time breakdown includes the execution time of FFT, point-point multiplication, Global Communication (Global Comm) phase, SYEVD, etc. 
The NDFT design includes the additional \textit{scheduling} overhead between CPU and NDP, as mentioned in Section~\ref{sec:worklod_scheduling}. 
We compare NDFT with both GPU and CPU baselines, and obtain the following evaluation results: 

\begin{figure}[htbp]
    \centering
    \vspace{-3mm}
    \includegraphics[width=\linewidth]{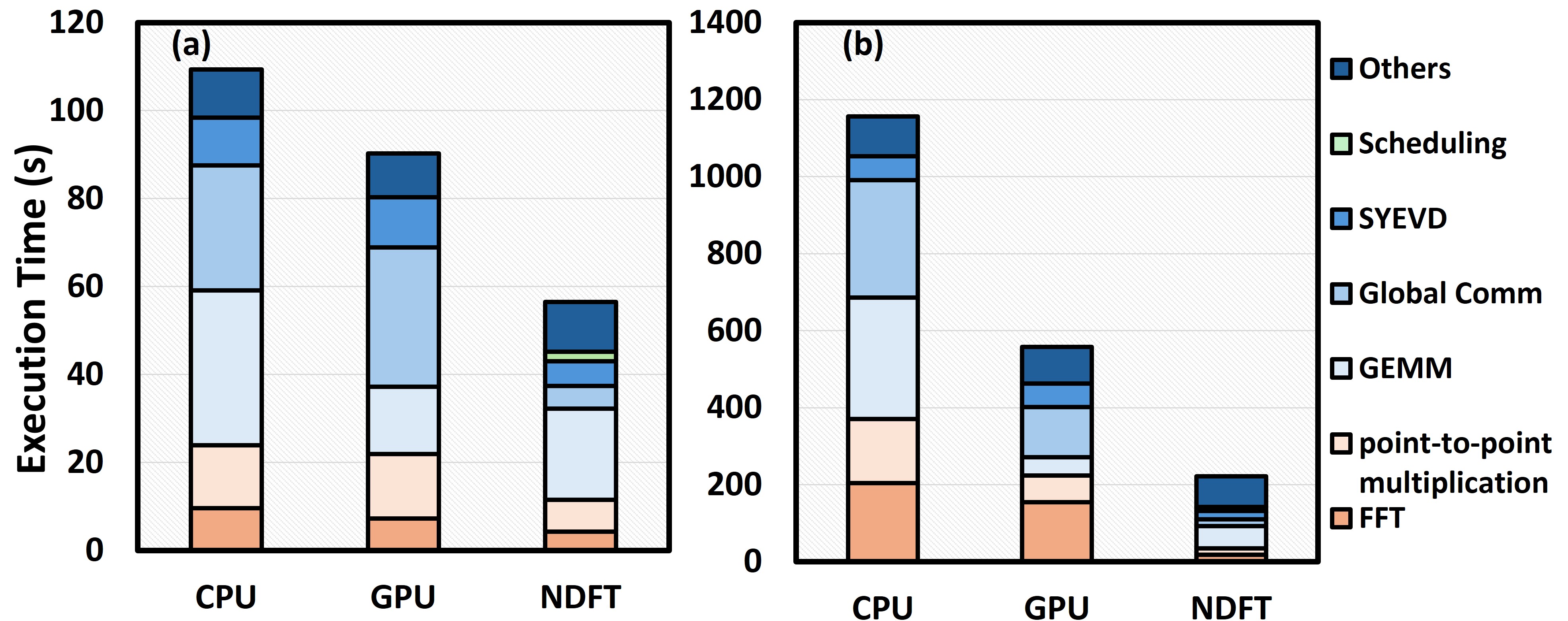}\\
    \caption{Performance comparison between CPU, GPU, and NDFT design using a small physical system (a) and a large physical system (b).}
    \label{fig:execution-time}
    \vspace{-3mm}
\end{figure}

\textbf{Comparison with CPU.}
(1) Despite the scheduling overhead, NDFT achieves 1.9x and 5.2x speedups in small and large physical systems, respectively, thanks to our scheduling mechanism and software/hardware co-design.
(2) The performance of memory-intensive kernels is significantly improved; for example, FFT achieves an 11.2x speedup in the large system, while the Face-splitting Product achieves a 1.99x speedup in the small system.

\textbf{Comparison with GPU.}
(1) NDFT achieves 1.6x and 2.5x speedups compared to the GPU baseline in small and large physical systems, respectively.
(2) While compute-bound kernels like GEMM on the GPU outperform those on NDFT by 35.9\% and 22.2\% in the two physical systems, respectively, memory-bound kernels show significant improvements with NDFT, achieving 2.1x and 5.2x speedups in the small and large physical systems, respectively.

\textbf{Other Discussion.}
(1) The scheduling overhead accounts for only 3.8\% and 4.9\% in small and large physical systems, respectively, demonstrating that NDFT's scheduling strategy carefully keeps this overhead minimal while improving kernel performance. 
(2) We study the memory footprint of pseudopotential information in NDFT. 
We observe that, in the large physical system, NDFT reduces memory footprint by 57.8\% compared to NDP in Table~\ref{tab:memory-usage}, bringing it close to CPU execution (1.08x). 
Global Comm only increases by 3.2\% over the GPU baseline, showing that NDFT’s hardware/software co-design resolves the OOM issue in NDP systems with low communication overhead. 

\subsection{Scalability Analysis}
\begin{figure}[htbp]
    \vspace{-3mm}
    \centering
    \includegraphics[width=\linewidth]{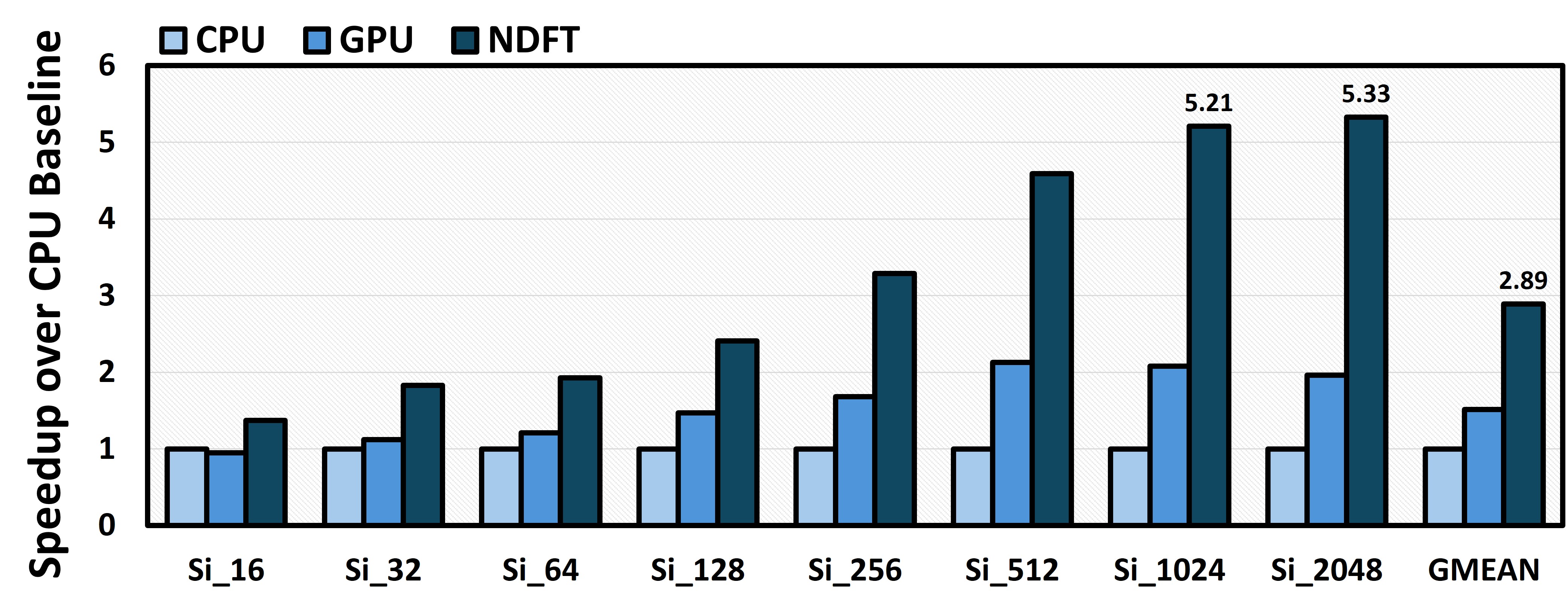}\\
    \caption{Speedup provided by NDFT, GPU over CPU baseline in several physical system scales.}
    \label{fig:scale}
    \vspace{-3mm}
\end{figure}
To explore NDFT’s performance across different physical system scales, we conduct a study ranging from Si\_16 to Si\_2048. 
Figure \ref{fig:scale} demonstrates that as the size of the physical system increases, NDFT’s performance advantage becomes more pronounced (up to 5.33x speedup in Si\_2048). 
We conclude that NDFT improves the performance of LR-TDDFT calculations in most cases, showing NDFT's great potential for addressing large-scale computational problems in the field of high-performance computing. 
\section{Related Work}
To our knowledge, NDFT is the first work to accelerate DFT calculations through optimizations of a general NDP architecture.
A prior work~\cite{liu2023heterogeneous} proposes a heterogeneous approach focused on alleviating data movement bottlenecks in quantum chemistry simulations based on \textit{ab initio} methods.
That work analyzes two specific components—FFT and time-consuming loops—proposing a custom hardware design tailored to these kernels.
In contrast, NDFT emphasizes task scheduling and hardware-software co-design, offering a more comprehensive solution within general-purpose NDP-CPU heterogeneous systems, thereby providing greater applicability and scalability.

\section{Conclusion}
In this work, we propose NDFT, a Near-Data Density Functional Theory framework to address performance bottlenecks in LR-TDDFT calculations on CPU-NDP heterogeneous systems.
NDFT involves a task partitioning and scheduling mechanism that efficiently assigns memory- and compute-bound kernels, minimizing data movement and fully leveraging CPU-NDP heterogeneous architectures, 
and a hardware-software co-design optimizing memory access for pseudopotential calculations, solving OOM issues in NDP systems. 
Evaluations show speedups of 5.2x and 2.5x over CPU and GPU baselines in a large physical system, as well as NDFT’s scalability across different system sizes. 


\section*{Acknowledgment}
This work is partly supported by Strategic Priority Research Program of the Chinese Academy of Sciences (Grant No. XDB0500102). The computing resources are financially supported by Laoshan Laboratory (LSKJ202300305).

\bibliographystyle{abbrv}
\bibliography{conference_101719}

\end{document}